\documentclass[aps,pra,twocolumn,reprint,nofootinbib,floatfix,superscriptaddress]{revtex4}
\usepackage{graphicx,bm,bbm,amsmath}
\usepackage{amssymb}
\usepackage[pdftex]{color} 
\usepackage{hyperref}
\usepackage{color}
\usepackage{ulem} 
\usepackage[paperwidth=8.5in,paperheight=11in,centering,hmargin=2cm,vmargin=2.5cm]{geometry} 

\newcommand{\bs}[1]{{\boldsymbol{#1}}}

\renewcommand{\b}{\bm}

\begin{document}

\title{Dissipationless conductance in a topological coaxial cable}

\author{Thomas~Schuster} \affiliation{Physics Department, Boston
 University, Boston, Massachusetts 02215, USA}

\author{Thomas~Iadecola} \affiliation{Physics Department, Boston
  University, Boston, Massachusetts 02215, USA}

\author{Claudio~Chamon} \affiliation{Physics Department, Boston
  University, Boston, Massachusetts 02215, USA}
  
\author{Roman~Jackiw} \affiliation{Department of Physics, 
Massachusetts Institute of Technology, Cambridge, Massachusetts 02139, USA}
  
\author{So-Young Pi} \affiliation{Physics Department, Boston
  University, Boston, Massachusetts 02215, USA}
  
\date{\today}

\begin{abstract}
We present a dynamical mechanism
leading to dissipationless conductance, whose quantized value is controllable,
in a (3+1)-dimensional electronic system.  The mechanism is exemplified by a theory of Weyl fermions
coupled to a Higgs field---also known as an axion insulator.  We show that the insertion of an axial gauge
flux can induce vortex lines in the Higgs field, similarly to the development of vortices
in a superconductor upon the insertion of
magnetic flux.  We further show that the necessary axial gauge flux can be
generated using Rashba spin-orbit coupling or a magnetic field.
Vortex lines in the Higgs field
are known to bind chiral fermionic modes, each of which serves as
a one-way channel for electric
charge with conductance $e^2/h$.  Combining these elements,
we present a physical picture, the ``topological coaxial cable,"  illustrating how 
the value of the quantized conductance could be controlled in such an axion insulator.
\end{abstract}

\maketitle

Quantized topological charges play an important role in the description
of quantum condensed matter systems with robust universal properties.
One of the most famous examples of this role occurs in the integer 
quantum Hall effect (IQHE)~\cite{VonKlitzing}, where the quantized Hall response 
is related to a topological invariant, the first Chern number~\cite{TKNN}.
Because the Chern number is quantized for topological reasons, the
quantized Hall response in the IQHE is robust to disorder, weak interactions,
and other perturbations, so long as they do not close the gap.  Variants of
the first Chern number have been used to demonstrate a similar level of
``topological protection" in the quantum spin Hall effect~\cite{KaneMele1,KaneMele2,BHZ}
and in topological semimetals~\cite{Wan11,BurkovBalents,BurkovHookBalents}.

Quantization also arises \textit{dynamically} in the context of flux quantization in type-II
superconductors.  In this case, inserting magnetic flux into the superconductor forces
the superconducting order parameter to acquire vorticity in order to screen the flux and
minimize the total energy~\cite{GinzburgLandau,Abrikosov}.  However, because
the order parameter must be single-valued over any closed path, its vorticity $n$ must be
quantized to integer values.  Thus, the superconducting order parameter screens the magnetic
flux one ``flux quantum" at a time, its phase winding by $2\pi\, n$ along a path that encircles $n$
magnetic flux quanta.

A third kind of quantization can occur in systems of Dirac fermions,
where index theorems (see, e.g., \cite{AtiyahSinger} and \cite{Weinberg}) guarantee the existence
of a conserved integer number of zero-energy solutions to the Dirac equation, provided
that the energy spectrum is symmetric about zero.  In systems where Dirac fermions
couple to a bosonic ``Higgs" field, or order parameter, index theorems are responsible
for the protection of isolated zero-energy modes that bind to topological defects in the
order parameter~\cite{JackiwRebbi,SSH,JackiwRossi,ReadGreen,Hou}.  
In two dimensions, the topological index that counts the number of zero modes associated
with a vortex in the order parameter is equal to the vorticity~\cite{Weinberg}.  Thus, a vortex with vorticity $n$
traps $n$ (orthogonal) zero modes.

In this paper, we provide an example of a system where these three disparate notions become
intimately related. The system in question consists of Weyl fermions in 3+1 spacetime dimensions, coupled
to a complex scalar field $\Delta$, which plays the role of an order parameter.  The field $\Delta$
acts as a Higgs field, in that the Weyl fermions acquire a mass when $\Delta$ develops a nonzero
expectation value.  Vortex \textit{lines} in $\Delta$ bind chiral fermionic modes
that are natural generalizations of the zero modes that bind to pointlike vortices in 2+1 dimensions~\cite{JackiwRossi,Witten85}.
These chiral fermionic modes serve as dissipationless conducting channels, each with
conductance $e^2/h$.  We demonstrate that a \textit{dynamical} treatment of the field $\Delta$
unveils a mechanism for the generation of such vortex lines---the insertion of a quantum of \textit{axial}
flux forces $\Delta$ to acquire a vortex profile in order to minimize the energy, in much the
same way that the order parameter in a type-II superconductor acquires a vortex profile upon the insertion
of a conventional magnetic flux.  Furthermore, we demonstrate that the axial flux necessary 
for the formation of the conducting channels can be generated in physically realizable ways,
either by the electric field associated with a line of charge (via the
Rashba effect) or by the magnetic field associated with a line of current (via the usual Pauli coupling).

This motivates us to consider a device, the ``topological coaxial cable," in which all of these
elements are present.  The device consists of a hollow cylinder of the insulating Weyl
system described above, with a line of charge or current at its center generating an
electric or magnetic field, and the corresponding axial vector potential.  Depending on the value of
the charge or current density,
an integer number of chiral fermionic modes, each of which serves as a dissipationless
conducting channel, nucleates around the inner surface of the cylinder. In this way, we demonstrate a novel 
dynamical means of producing quantized dissipationless conductance in a (3+1)-dimensional electronic system.

We begin with a Weyl semimetal with two zero-energy Weyl points, at momenta $\b{K_{\pm}} = \pm k_0 \b{\hat{z}}$. We further suppose there exists a dynamical complex scalar (Higgs) field $\Delta(\b{r})$~\cite{Hou,WangZhang} that couples the two Weyl points and an axial gauge potential $\bm{A}_5 = (A_{5,x},A_{5,y},A_{5,z})$ coupling to the fermions. The Hamiltonian of the system is $H = \int d^3\b{r} \, \Psi^{\dagger}(\b{r}) \mathcal{H}(\b{r}) \Psi(\b{r})$, where $\Psi(\b{r})$ is a 4-component spinor $\Psi^{\dagger} = \big( \psi_{+,\uparrow}, \psi_{+,\downarrow}, \psi_{-,\uparrow}, -\psi_{-,\downarrow} \big)$ where $+/-$ indices label each Weyl point and $\uparrow/\downarrow$ label the electron spin, and $\mathcal{H}(\b{r})$ is the $4 \times 4$ matrix
\begin{equation} \label{Hamiltonian matrix}
\mathcal{H}(\b{r})  = 
\begin{pmatrix}
(-i \bm{\partial} - \bm{A}_5)  \cdot \bm{\sigma} & \Delta(\b{r}) \mathbbm{1} \\
\Delta^*(\b{r}) \mathbbm{1} & (i \bm{\partial} - \bm{A}_5) \cdot \bm{\sigma}
\end{pmatrix},
\end{equation}
with $\bm{\partial} = (\partial_x,\partial_y,\partial_z)$ and the Pauli matrices $\bm{\sigma} = (\sigma_x,\sigma_y, \sigma_z)$.  In the absence of the axial gauge potential and the Higgs field, Eq.~\eqref{Hamiltonian matrix} describes the low-energy limit of the tight-binding model presented in Ref.~\cite{YangLuRan}. Although we have assumed a particular form of spin-momentum locking at the Weyl points in writing down this Hamiltonian, we stress that what follows is not dependent on the details of the spin-momentum locking (with one exception, mentioned later on). What is necessary is a Weyl semimetal that breaks time-reversal symmetry, such that the field $\Delta$ may couple a single pair of Weyl points with opposite chirality.

To analyze the fermion dynamics in this Hamiltonian, we first set $\bm{A}_5 = 0$ and consider a fixed (\textit{i.e.} not dynamical) $\Delta(\b{r})$. For a constant $\Delta$, a band gap of magnitude $2|\Delta|$ is opened at both Weyl points, and the Weyl semimetal becomes an ``axion insulator"~\cite{WangZhang}. We are interested in the case where $\Delta$ has a constant magnitude but hosts a vortex line (here taken to extend in the $z$-direction), with a twist of vorticity $n$ in its phase,
 \begin{equation}\label{Delta vortex}
 \Delta(\b{r}) = \Delta_0(r) e^{i n \theta},
 \end{equation} 
 working in cylindrical coordinates $\bm r = (r,\theta,z) = (\sqrt{x^2+y^2},\tan^{-1}(y/x),z)$. The magnitude of the field, $\Delta_0(r)$, is $0$ at the vortex center (which we place at the origin) and approaches some constant value $\Delta_0$ far from the vortex. To solve for the fermion modes, first note that we have retained translational invariance in the $z$-direction, and can thus take the eigenvalue of $-i\partial_z$, $p_z$, as a good quantum number. In the special case $p_z = 0$, the Hamiltonian is identical to that of graphene with a Kekul\'{e} vortex, which features $n$ (pseudo)spin-polarized zero modes exponentially localized at the vortex center~\cite{Hou}. When $n=1$ (see Appendix for wavefunctions with arbitrary vorticity) and for nonzero $p_z$, the vortex mode wavefunctions are
\begin{equation}\label{vortex mode}
\Psi^v_{p_z}(\b{r})
=
e^{i p_z z} e^{-\int_0^r dr' \, \Delta_0(r')}
\begin{pmatrix} 
 e^{i\pi/4} \\ 
 0  \\
0  \\
 e^{-i\pi/4} \\
\end{pmatrix},
\, E = p_z.
\end{equation}
Thus we have a single fermionic mode, localized at the vortex center, with a linear, chiral dispersion relation. This result extends to higher vorticities: for a vortex of vorticity $n$, there are $n$ linearly-independent vortex modes, each with the same chiral dispersion relation. Furthermore, in a finite-size system, there are $n$ modes of the opposite chirality (i.e., whose dispersion is given by $E = -p_z$) exponentially localized at the outer boundary. These chiral fermionic modes are the origin of the quantized conductance---since the Weyl semimetal becomes insulating when the Higgs field acquires an expectation value, the vortex and boundary modes are the sole current-carriers at low temperature~\cite{FermiArcFootnote}. Current flowing in the positive $z$-direction will be carried by the vortex modes, while current flowing in the negative $z$-direction will be carried along the outer boundary. Provided the vortex and boundary are sufficiently distant, these modes are immune to backscattering, and thus lead to a quantized conductance $\sigma = ne^2/h$ where $n$ is the number of vortex modes.  

Similar chiral modes bound to vortex lines in 3+1 dimensions were considered in the context of cosmic strings~\cite{Witten85,CallanHarvey}, and more recently in topological superconductors, Weyl semimetals, and axion insulators~\cite{WangZhang,QiWittenZhang, LiuYeQi,StoneLopes}.  In the condensed-matter context, the vortex lines hosting the chiral fermionic modes have typically been inserted by hand, leaving their physical origin unspecified.  In Ref.~\cite{WangZhang}, it was pointed out that dislocations in a charge density wave, which can appear during thermal annealing of the axion insulator, can also form such vortex lines.  However, taking advantage of the quantized conductance $ne^2/h$ becomes problematic in this case, since the locations and density of the dislocations are probabilistic in nature and vary from system to system.  In the remainder of this paper, we describe a \textit{dynamical} mechanism by which vortex lines can be formed in a controlled manner in the axion insulator described by the Hamiltonian~\eqref{Hamiltonian matrix}.  An added benefit of the mechanism is that the winding number $n$ of the vortex line can also be controlled externally, so that the quantized conductance is tunable.

We now show that a certain configuration of an externally applied axial gauge potential will induce a vortex in the dynamical field $\Delta$. To see this, we integrate out the fermions and consider the effective Hamiltonian for $\Delta$ in the presence of an external axial gauge field. The form of this Hamiltonian is fixed by gauge invariance, and reads (see also~\cite{SSH})
\begin{equation}\label{H eff}
\mathcal{H}_{\text{eff}}[\Delta(\b{r})] = | (\bm{\partial} - 2 i \bm{A}_5(\b{r})) \, \Delta(\b{r}) |^{2} + \mathcal{V}_{\text{eff}}[\Delta(\b{r})].
\end{equation}
We assume that the effective potential $\mathcal{V}_{\text{eff}}[\Delta(\b{r})]$ takes the form of a ``Mexican hat", so that the magnitude of $\Delta$ acquires a background expectation value $|\langle \Delta(\b{r}) \rangle| = \Delta_0$ (c.f.~Ref.~\cite{WangZhang}). We leave the phase of $\Delta$ to vary so that, in the presence of a fixed background profile of the axial gauge potential $\bm{A}_5$, it can adjust itself in order minimize the effective energy \eqref{H eff}. Indeed, a vortex line of vorticity $n$ in $\Delta$ minimizes the energy if the axial gauge potential takes the form 
\begin{align}\label{axial vortex}
\begin{split}
A_{5,i}(\b{r}) = -n \, \epsilon_{ij} & \frac{r_j}{2 \, r^2}, \, i,j = x,y,  \\
A_{5,z}(\b{r}) & = 0.
\end{split}
\end{align}
In other words, a vortex line of vorticity $n$ in $\Delta$ arises in the presence of a background profile~\eqref{axial vortex} in the axial gauge potential $\bm{A}_5$. Such an axial gauge potential was introduced for graphene in Ref.~\cite{JackiwPi}.
It is worth noting that the vortex and edge modes persist in the presence of $\bm{A}_5$ for topological reasons (see also~\cite{JackiwPi}). The Hamiltonian (\ref{Hamiltonian matrix}) falls into Altland-Zirnbauer symmetry class A, which is capable of supporting codimension-2 topological defects with a $\mathbbm{Z}$ classification~\cite{TeoKane} (in our case, the vorticity). 




We now address the question of how to generate an axial gauge potential of the form \eqref{axial vortex} in a particular physical realization of the Hamiltonian (\ref{Hamiltonian matrix}). For a physical realization whose low-energy theory is described by Eq.~\eqref{Hamiltonian matrix} in the spinor basis $\Psi^{\dagger} = \big( \psi_{+,\uparrow}, \psi_{+,\downarrow}, \psi_{-,\uparrow}, -\psi_{-,\downarrow} \big)$, the external axial gauge field (\ref{axial vortex}) can be induced by an ordinary electric field through Rashba spin-orbit coupling (RSOC). RSOC describes the coupling of the spin of an electron moving in an electric field to the magnetic field seen in that electron's rest frame. Namely, $H_R = \alpha_R (\b{q} \times \b{E} ) \cdot \bm{\sigma}$ where $\b{q}$ is the electron momentum and $\alpha_R$ is some constant. To zeroth order in the quasiparticle momentum $\b{p}$, we can simply set $\b{q} = \b{K}_{\pm} = \pm k_0 \b{\hat{z}}$ for each Weyl point, and find that the RSOC enters the low-energy theory as 
\begin{equation}\label{H rashba}
\mathcal{H}_{R} =  \alpha_R' 
\begin{pmatrix} 
(k_0 \b{\hat{z}} \times \b{E} ) \cdot \bm{\sigma} & 0 \\ 
0 & (k_0 \b{\hat{z}} \times \b{E} ) \cdot \bm{\sigma}  \\
\end{pmatrix},
\end{equation} 
where $\alpha_R'$ differs from $\alpha_R$ due to lattice effects. Comparing to the Hamiltonian (\ref{Hamiltonian matrix}), we see this is just a 2D axial gauge field with $A_{5,i} = \alpha_R' k_0 \, \epsilon_{ij} E_j, \, i,j = x,y$, and $A_{5,z} = 0$. A vortex (\ref{axial vortex}) in the axial gauge field corresponds to $E_{i}(\b{r}) = \frac{-n}{\alpha_R' k_0} \frac{r_i}{2 \, r^2}$, which is just the electric field generated by a stationary wire extending in the $z$-direction, with linear charge density $\lambda = \frac{-n \pi \epsilon}{\alpha_R'}$.

Although RSOC entered the low energy limit of the specific Hamiltonian (\ref{Hamiltonian matrix}) as an axial gauge field, this would not be true for a different pattern of spin-momentum locking. For example, consider a physical realization whose Hamiltonian has the same form as Eq. (\ref{Hamiltonian matrix}) at low energies, but for which the natural spinor basis is $\Psi^{\dagger} = \big( \psi_{+,\uparrow}, \psi_{+,\downarrow}, \psi_{-,\uparrow}, \psi_{-,\downarrow} \big)$ (this would occur in a physical system respecting parity symmetry).  In this case, RSOC would enter as an ordinary gauge potential, and we need to look elsewhere for the desired axial gauge potential. However, it turns out that the Pauli coupling of the electron's spin to an external \textit{magnetic} field, $H_B = \alpha_B \bm{B} \cdot \bs{\sigma}$ now enters as an axial gauge potential,
\begin{equation}\label{H magnetic}
\mathcal{H}_{B} =  \alpha_B 
\begin{pmatrix} 
\bm{B} \cdot \bm{\sigma} & 0 \\ 
0 & \bm{B} \cdot \bm{\sigma}  \\
\end{pmatrix},
\end{equation} 
with $A_{5,i} = -\alpha_B B_i, \, i = x,y,z$. A vortex in $\bm{A}_5$ now corresponds to the magnetic field of a current-carrying wire along the vortex line, of current $I = \frac{-n \pi}{\alpha_B \mu}$. Clearly a continuum of spin-momentum locking scenarios at each Weyl point exist; with these examples we mean only to highlight that an external axial gauge potential is not unphysical.

\begin{figure}
\centering
\includegraphics[width=.3\textwidth,page=1]{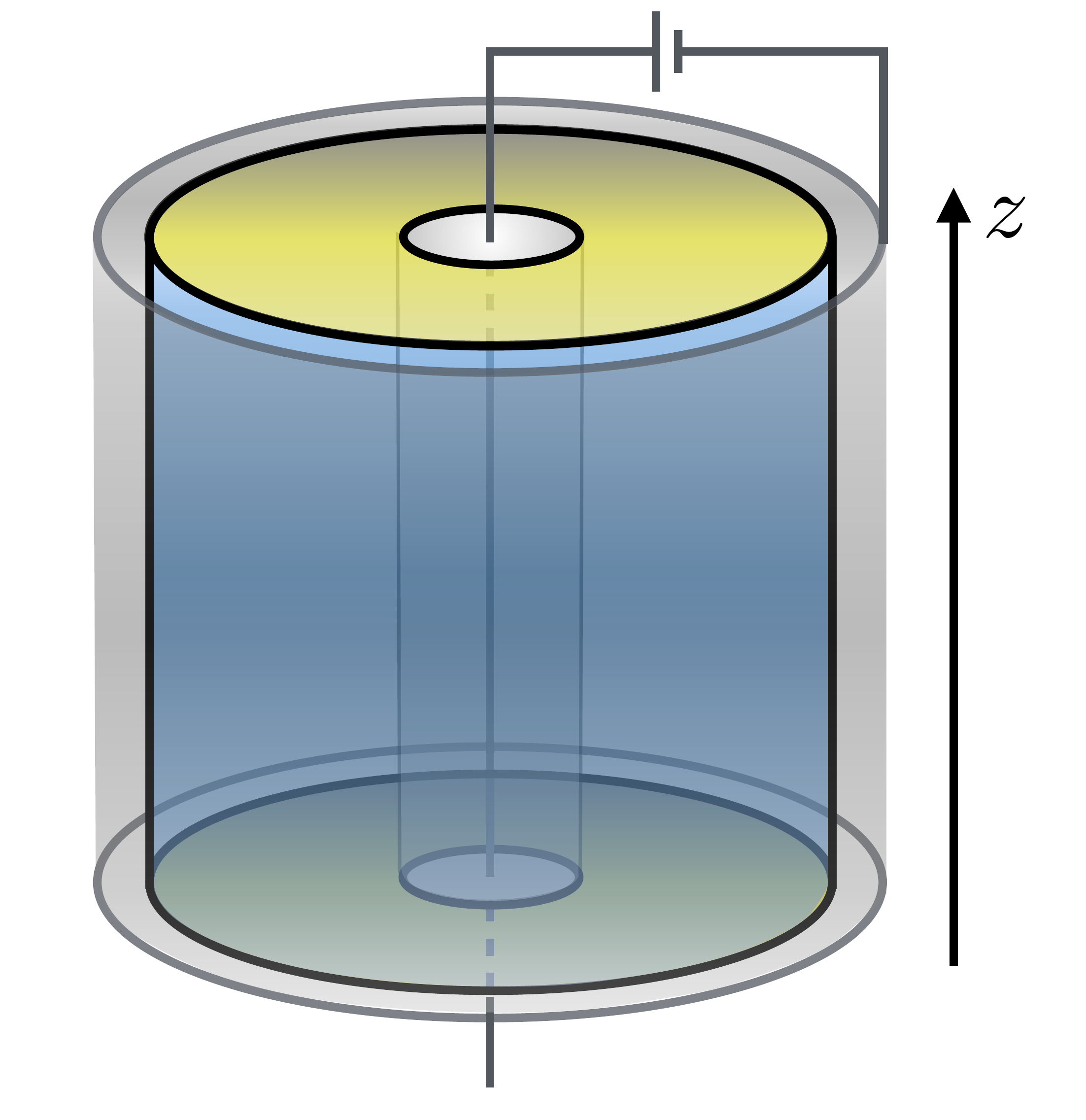}
\caption{Schematic of the topological coaxial cable geometry. The axion insulator (blue) is enclosed in a cylindrical capacitor (silver). A voltage applied to the capacitor creates the axial gauge field vortex (\ref{axial vortex}) in the axion insulator through RSOC. Annular contacts (gold) are placed on both ends to measure the longitudinal conductance, allowing current to be carried through the vortex and edge modes. The number of such modes is dependent on the voltage across the capacitor. \label{fig: schematic}}
\end{figure}
%
%
%

\begin{figure}
\centering
\includegraphics[width=.475\textwidth,page=2]{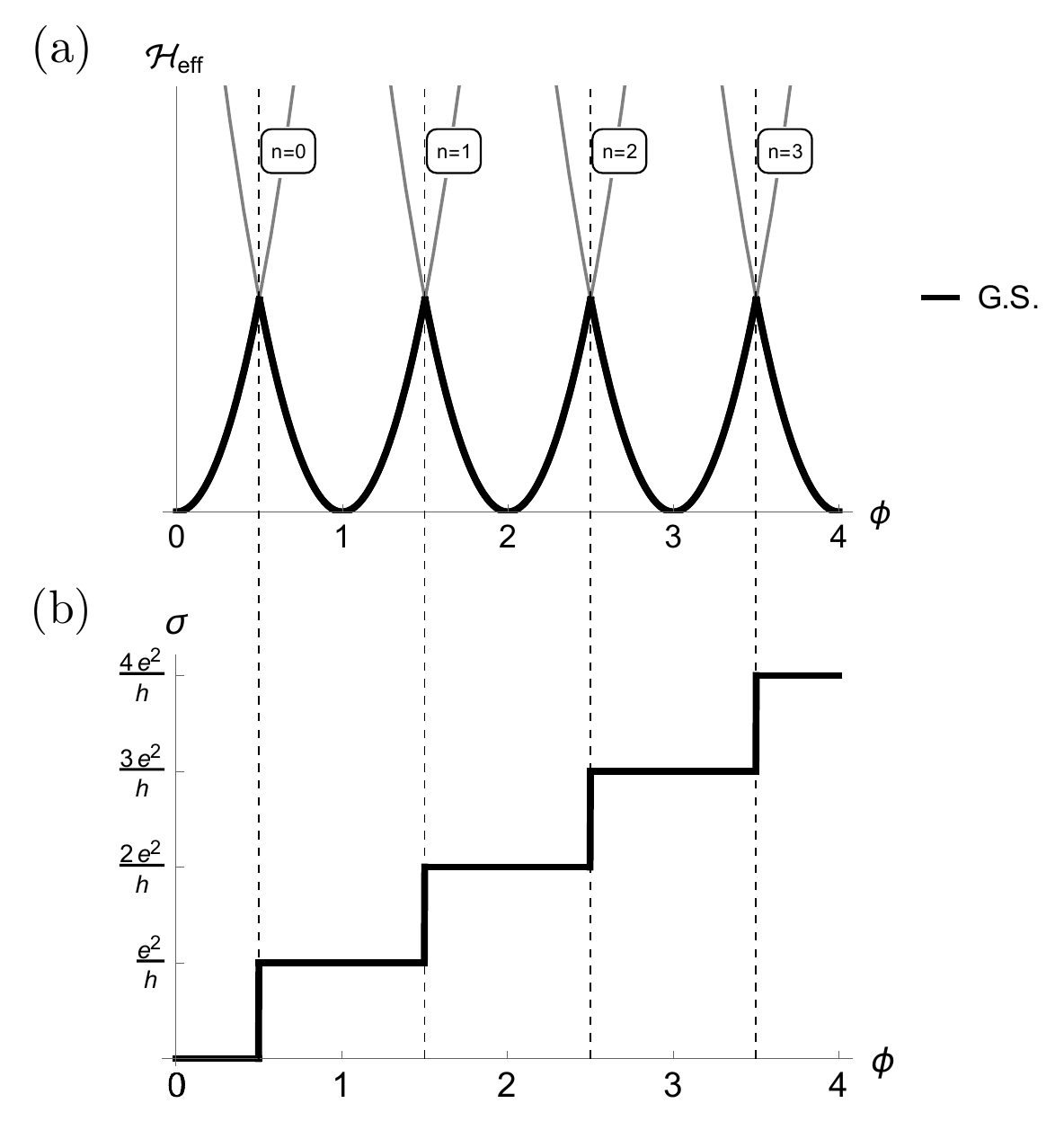}
\caption{(a) Energy of the effective Hamiltonian plotted against the applied axial gauge field strength $\phi$ for the lowest few vorticities $n$ of $\Delta$, as given by Eq. (\ref{H eff energy}). This energy is minimized when $n$ is the nearest integer to $\phi$ (the thick black line). (b) Longitudinal conductivity $\sigma$ of the topological coaxial cable plotted against the applied axial gauge field strength $\phi$. The conductivity is a step-function in $\phi$, with jumps when $\phi$ takes half-integer values, corresponding to changes in the vorticity $n$ which minimizes the effective Hamiltonian. Dashed lines highlight this correspondence. \label{fig: phi}}
\end{figure}

Motivated by this, and choosing to work with the RSOC instance of an axial gauge potential in (\ref{H rashba}), we consider the following ``topological coaxial cable'' (TCC) geometry: a cylindrical capacitor of outer radius $R$ and inner radius $a$, filled with the axion insulator (\ref{Hamiltonian matrix}) in place of a dielectric (see Fig.~\ref{fig: schematic}). A voltage $V$ applied to the capacitor creates an electric field $E_i(\b{r}) = \lambda r_i/2 \pi \epsilon \, r^2$, with $\lambda = 2 \pi \epsilon V/ \ln(R/a)$, in the axion insulator, and thus an axial gauge potential $A_{5,i}(\b{r}) = -\phi \, \epsilon_{ij} \frac{r_j}{2 \, r^2}$, where $\phi \equiv  \alpha_R' \lambda / \pi \epsilon$. Note that when $\phi$ is an integer it corresponds to $n$ in Eq.~(\ref{axial vortex}), which favors a vortex in $\Delta$ of vorticity $n$. For noninteger values of $\phi$ the covariant derivative in the effective Hamiltonian (\ref{H eff}) cannot be made to vanish due to the quantization of vorticity in $\Delta$. We make the ansatz that the covariant derivative is minimized if $\Delta$ takes a form described by Eq.~(\ref{Delta vortex}), with $n$ the nearest integer to $\phi$. This configuration has energy 
\begin{align}\label{H eff energy}
\begin{split}
H_{\text{eff}} & = \int d^3 \b{r} | (\bm{\partial} - 2 i \bm{A}_5(\b{r})) \Delta(\b{r}) |^{2} \\
& = (n-\phi)^2 \Delta_0^2 \int d^3 \b{r} \, \frac{1}{r^2}.
\end{split}
\end{align}
This energy for each $n$, as well as the ground state energy (where $n$ is taken to be the nearest integer to $\phi$), is plotted in Fig. \ref{fig: phi} (a).

We now consider the current carried by the vortex modes in the TCC when the voltage $V$ is fixed such that the quantity $\phi=n$, thus minimizing $H_{\text{eff}}$.  This is given by the famous Callan-Harvey formula for the current density~\cite{CallanHarvey},
\begin{align}
j^{\mu}=\frac{1}{4\pi}\frac{e^2}{h}\epsilon^{\mu\nu\rho\sigma}\, \partial_\nu\varphi\, F_{\rho\sigma},
\end{align}
where $\varphi\equiv \text{arg }\Delta$ and $F_{\mu\nu}$ is the Maxwell field strength tensor describing the probe electromagnetic gauge potential.  When $\phi=n$, we have $\varphi = n\, \theta$ [c.f.~Eq.~\eqref{Delta vortex}], so the density of current carried in the $z$-direction is
\begin{align}\label{z current density}
j^{z}=-n\frac{e^2}{h}\, \frac{E_r}{2\pi},
\end{align}
where $E_r=F_{0r}$ is the radial component of the applied electric field.  From Eq.~\eqref{z current density}, we can read off the conductance $\sigma = n e^2/h$ directly.  Furthermore, we see that a \textit{radial} electric field drives a \textit{vertical} current.  Thus, in the RSOC realization of a TCC, the radial electric field that induces the axial vortex also acts as a source of dissipationless current in the $z$-direction.

When the applied voltage $V$ is detuned from the special values such that $\phi=n$, the longitudinal conductance becomes $\sigma = \lfloor\phi\rceil\, e^2/h$, where $\lfloor\phi\rceil$ is the nearest integer to $\phi =  \alpha_R' \lambda / \pi \epsilon$ [see Fig. \ref{fig: phi} (b)]. Thus although the capacitor charge density $\lambda$ varies continuously, the conductance is quantized as a step function in $\lambda$, owing to the quantization of vorticity in $\Delta$. This conductance will saturate when the outermost chiral vortex modes have significant overlap with the corresponding edge modes of opposite chirality, in which case these modes will gap out and cease to add to the conductivity. Also note that, were we to alter this set-up to work with the Pauli coupling instance of an axial gauge field (\ref{H magnetic}), our results would be identical after the replacement $\lambda \rightarrow \frac{\alpha_B \epsilon \mu I}{\alpha_R}$.  (In this case, one must additionally apply a radial electric field in order to drive current in the $z$-direction.)

In summary, we have demonstrated a dynamical mechanism for the development of quantized conductivity in axion insulators that descend from time-reversal breaking Weyl semimetals upon the introduction of a Higgs field (or order parameter). Vortices in the Higgs field bind chiral, linearly dispersing modes, which are the sole bulk current carriers at low temperatures. We showed that such vortices can be induced dynamically by the insertion of an axial gauge flux---in the presence of such an axial flux, the phase of the order parameter acquires a twist so as to minimize energy.  The magnitude of the applied gauge potential controls the vorticity of the induced twist, and therefore also sets the number of chiral conducting channels. These features come together in the ``topological coaxial cable", in which the axial gauge potential arises from Rashba spin-orbit coupling (or, in another physical realization, from the Pauli coupling of an applied magnetic field to the electron's spin).  The coaxial cable carries current along its axis in response to an applied radial electric field, and features a dissipationless conductance $\sigma = ne^2/h$, where $n$ is the number of induced chiral channels.

\begin{acknowledgments}
We thank Luiz H. Santos for helpful feedback on the manuscript. T.I. was supported by the National Science Foundation Graduate Research Fellowship Program under Grant No.~DGE-1247312, and C.C. was supported by DOE Grant DEF-06ER46316.
\end{acknowledgments}

\bibliographystyle{apsrev}

\bibliography{refs_coaxial}

\begin{thebibliography}{27}
\expandafter\ifx\csname natexlab\endcsname\relax\def\natexlab#1{#1}\fi
\expandafter\ifx\csname bibnamefont\endcsname\relax
  \def\bibnamefont#1{#1}\fi
\expandafter\ifx\csname bibfnamefont\endcsname\relax
  \def\bibfnamefont#1{#1}\fi
\expandafter\ifx\csname citenamefont\endcsname\relax
  \def\citenamefont#1{#1}\fi
\expandafter\ifx\csname url\endcsname\relax
  \def\url#1{\texttt{#1}}\fi
\expandafter\ifx\csname urlprefix\endcsname\relax\def\urlprefix{URL }\fi
\providecommand{\bibinfo}[2]{#2}
\providecommand{\eprint}[2][]{\url{#2}}

\bibitem[{\citenamefont{Klitzing et~al.}(1980)\citenamefont{Klitzing, Dorda,
  and Pepper}}]{VonKlitzing}
\bibinfo{author}{\bibfnamefont{K.~v.} \bibnamefont{Klitzing}},
  \bibinfo{author}{\bibfnamefont{G.}~\bibnamefont{Dorda}}, \bibnamefont{and}
  \bibinfo{author}{\bibfnamefont{M.}~\bibnamefont{Pepper}},
  \bibinfo{journal}{Phys. Rev. Lett.} \textbf{\bibinfo{volume}{45}},
  \bibinfo{pages}{494} (\bibinfo{year}{1980}).

\bibitem[{\citenamefont{Thouless et~al.}(1982)\citenamefont{Thouless, Kohmoto,
  Nightingale, and den Nijs}}]{TKNN}
\bibinfo{author}{\bibfnamefont{D.~J.} \bibnamefont{Thouless}},
  \bibinfo{author}{\bibfnamefont{M.}~\bibnamefont{Kohmoto}},
  \bibinfo{author}{\bibfnamefont{M.~P.} \bibnamefont{Nightingale}},
  \bibnamefont{and} \bibinfo{author}{\bibfnamefont{M.}~\bibnamefont{den Nijs}},
  \bibinfo{journal}{Phys. Rev. Lett.} \textbf{\bibinfo{volume}{49}},
  \bibinfo{pages}{405} (\bibinfo{year}{1982}).

\bibitem[{\citenamefont{Kane and Mele}(2005{\natexlab{a}})}]{KaneMele1}
\bibinfo{author}{\bibfnamefont{C.~L.} \bibnamefont{Kane}} \bibnamefont{and}
  \bibinfo{author}{\bibfnamefont{E.~J.} \bibnamefont{Mele}},
  \bibinfo{journal}{Phys. Rev. Lett.} \textbf{\bibinfo{volume}{95}},
  \bibinfo{pages}{146802} (\bibinfo{year}{2005}{\natexlab{a}}).

\bibitem[{\citenamefont{Kane and Mele}(2005{\natexlab{b}})}]{KaneMele2}
\bibinfo{author}{\bibfnamefont{C.~L.} \bibnamefont{Kane}} \bibnamefont{and}
  \bibinfo{author}{\bibfnamefont{E.~J.} \bibnamefont{Mele}},
  \bibinfo{journal}{Phys. Rev. Lett.} \textbf{\bibinfo{volume}{95}},
  \bibinfo{pages}{226801} (\bibinfo{year}{2005}{\natexlab{b}}).

\bibitem[{\citenamefont{Bernevig et~al.}(2006)\citenamefont{Bernevig, Hughes,
  and Zhang}}]{BHZ}
\bibinfo{author}{\bibfnamefont{B.~A.} \bibnamefont{Bernevig}},
  \bibinfo{author}{\bibfnamefont{T.~L.} \bibnamefont{Hughes}},
  \bibnamefont{and} \bibinfo{author}{\bibfnamefont{S.-C.} \bibnamefont{Zhang}},
  \bibinfo{journal}{Science} \textbf{\bibinfo{volume}{314}},
  \bibinfo{pages}{1757} (\bibinfo{year}{2006}).

\bibitem[{\citenamefont{Wan et~al.}(2011)\citenamefont{Wan, Turner, Vishwanath,
  and Savrasov}}]{Wan11}
\bibinfo{author}{\bibfnamefont{X.}~\bibnamefont{Wan}},
  \bibinfo{author}{\bibfnamefont{A.~M.} \bibnamefont{Turner}},
  \bibinfo{author}{\bibfnamefont{A.}~\bibnamefont{Vishwanath}},
  \bibnamefont{and} \bibinfo{author}{\bibfnamefont{S.~Y.}
  \bibnamefont{Savrasov}}, \bibinfo{journal}{Phys. Rev. B}
  \textbf{\bibinfo{volume}{83}}, \bibinfo{pages}{205101}
  (\bibinfo{year}{2011}).

\bibitem[{\citenamefont{Burkov and Balents}(2011)}]{BurkovBalents}
\bibinfo{author}{\bibfnamefont{A.~A.} \bibnamefont{Burkov}} \bibnamefont{and}
  \bibinfo{author}{\bibfnamefont{L.}~\bibnamefont{Balents}},
  \bibinfo{journal}{Phys. Rev. Lett.} \textbf{\bibinfo{volume}{107}},
  \bibinfo{pages}{127205} (\bibinfo{year}{2011}).

\bibitem[{\citenamefont{Burkov et~al.}(2011)\citenamefont{Burkov, Hook, and
  Balents}}]{BurkovHookBalents}
\bibinfo{author}{\bibfnamefont{A.~A.} \bibnamefont{Burkov}},
  \bibinfo{author}{\bibfnamefont{M.~D.} \bibnamefont{Hook}}, \bibnamefont{and}
  \bibinfo{author}{\bibfnamefont{L.}~\bibnamefont{Balents}},
  \bibinfo{journal}{Phys. Rev. B} \textbf{\bibinfo{volume}{84}},
  \bibinfo{pages}{235126} (\bibinfo{year}{2011}).

\bibitem[{Gin()}]{GinzburgLandau}
\bibinfo{howpublished}{V.L. Ginzburg and L.D. Landau, Zh. Eksp. Teor. Fiz.
  \textbf{20}, 1064 (1950).}

\bibitem[{Abr()}]{Abrikosov}
\bibinfo{howpublished}{A.A. Abrikosov, Sov. Phys. JETP \textbf{5}, 1174
  (1957).}

\bibitem[{Ati()}]{AtiyahSinger}
\bibinfo{howpublished}{M. F. Atiyah and I. M. Singer, Bull. Amer. Math. Soc.
  \textbf{69}, 422 (1963).}

\bibitem[{\citenamefont{Weinberg}(1981)}]{Weinberg}
\bibinfo{author}{\bibfnamefont{E.~J.} \bibnamefont{Weinberg}},
  \bibinfo{journal}{Phys. Rev. D} \textbf{\bibinfo{volume}{24}},
  \bibinfo{pages}{2669} (\bibinfo{year}{1981}).

\bibitem[{\citenamefont{Jackiw and Rebbi}(1976)}]{JackiwRebbi}
\bibinfo{author}{\bibfnamefont{R.}~\bibnamefont{Jackiw}} \bibnamefont{and}
  \bibinfo{author}{\bibfnamefont{C.}~\bibnamefont{Rebbi}},
  \bibinfo{journal}{Phys. Rev. D} \textbf{\bibinfo{volume}{13}},
  \bibinfo{pages}{3398} (\bibinfo{year}{1976}).

\bibitem[{\citenamefont{Su et~al.}(1979)\citenamefont{Su, Schrieffer, and
  Heeger}}]{SSH}
\bibinfo{author}{\bibfnamefont{W.~P.} \bibnamefont{Su}},
  \bibinfo{author}{\bibfnamefont{J.~R.} \bibnamefont{Schrieffer}},
  \bibnamefont{and} \bibinfo{author}{\bibfnamefont{A.~J.}
  \bibnamefont{Heeger}}, \bibinfo{journal}{Phys. Rev. Lett.}
  \textbf{\bibinfo{volume}{42}}, \bibinfo{pages}{1698} (\bibinfo{year}{1979}).

\bibitem[{\citenamefont{Jackiw and Rossi}(1981)}]{JackiwRossi}
\bibinfo{author}{\bibfnamefont{R.}~\bibnamefont{Jackiw}} \bibnamefont{and}
  \bibinfo{author}{\bibfnamefont{P.}~\bibnamefont{Rossi}},
  \bibinfo{journal}{Nuclear Physics B} \textbf{\bibinfo{volume}{190}},
  \bibinfo{pages}{681 } (\bibinfo{year}{1981}).

\bibitem[{\citenamefont{Read and Green}(2000)}]{ReadGreen}
\bibinfo{author}{\bibfnamefont{N.}~\bibnamefont{Read}} \bibnamefont{and}
  \bibinfo{author}{\bibfnamefont{D.}~\bibnamefont{Green}},
  \bibinfo{journal}{Phys. Rev. B} \textbf{\bibinfo{volume}{61}},
  \bibinfo{pages}{10267} (\bibinfo{year}{2000}).

\bibitem[{\citenamefont{Hou et~al.}(2007)\citenamefont{Hou, Chamon, and
  Mudry}}]{Hou}
\bibinfo{author}{\bibfnamefont{C.-Y.} \bibnamefont{Hou}},
  \bibinfo{author}{\bibfnamefont{C.}~\bibnamefont{Chamon}}, \bibnamefont{and}
  \bibinfo{author}{\bibfnamefont{C.}~\bibnamefont{Mudry}},
  \bibinfo{journal}{Phys. Rev. Lett.} \textbf{\bibinfo{volume}{98}},
  \bibinfo{pages}{186809} (\bibinfo{year}{2007}).

\bibitem[{\citenamefont{Witten}(1985)}]{Witten85}
\bibinfo{author}{\bibfnamefont{E.}~\bibnamefont{Witten}},
  \bibinfo{journal}{Nucl. Phys. B} \textbf{\bibinfo{volume}{249}},
  \bibinfo{pages}{557} (\bibinfo{year}{1985}).

\bibitem[{\citenamefont{Wang and Zhang}(2013)}]{WangZhang}
\bibinfo{author}{\bibfnamefont{Z.}~\bibnamefont{Wang}} \bibnamefont{and}
  \bibinfo{author}{\bibfnamefont{S.-C.} \bibnamefont{Zhang}},
  \bibinfo{journal}{Phys. Rev. B} \textbf{\bibinfo{volume}{87}},
  \bibinfo{pages}{161107} (\bibinfo{year}{2013}).

\bibitem[{\citenamefont{Yang et~al.}(2011)\citenamefont{Yang, Lu, and
  Ran}}]{YangLuRan}
\bibinfo{author}{\bibfnamefont{K.-Y.} \bibnamefont{Yang}},
  \bibinfo{author}{\bibfnamefont{Y.-M.} \bibnamefont{Lu}}, \bibnamefont{and}
  \bibinfo{author}{\bibfnamefont{Y.}~\bibnamefont{Ran}},
  \bibinfo{journal}{Phys. Rev. B} \textbf{\bibinfo{volume}{84}},
  \bibinfo{pages}{075129} (\bibinfo{year}{2011}).

\bibitem[{Fer()}]{FermiArcFootnote}
\bibinfo{howpublished}{In a finite system, the axion insulator may also possess
  surface states that are remnants of the Fermi arc surface states of the
  parent Weyl semimetal. However, in a cylindrical geometry, these surface
  states can only carry azimuthal current, and thus do not contribute to
  transport in the $z$-direction, which is of interest to us here.}

\bibitem[{\citenamefont{Callan and Harvey}(1985)}]{CallanHarvey}
\bibinfo{author}{\bibfnamefont{C.}~\bibnamefont{Callan}} \bibnamefont{and}
  \bibinfo{author}{\bibfnamefont{J.}~\bibnamefont{Harvey}},
  \bibinfo{journal}{Nuclear Physics B} \textbf{\bibinfo{volume}{250}},
  \bibinfo{pages}{427 } (\bibinfo{year}{1985}).

\bibitem[{\citenamefont{Qi et~al.}(2013)\citenamefont{Qi, Witten, and
  Zhang}}]{QiWittenZhang}
\bibinfo{author}{\bibfnamefont{X.-L.} \bibnamefont{Qi}},
  \bibinfo{author}{\bibfnamefont{E.}~\bibnamefont{Witten}}, \bibnamefont{and}
  \bibinfo{author}{\bibfnamefont{S.-C.} \bibnamefont{Zhang}},
  \bibinfo{journal}{Phys. Rev. B} \textbf{\bibinfo{volume}{87}},
  \bibinfo{pages}{134519} (\bibinfo{year}{2013}).

\bibitem[{\citenamefont{Liu et~al.}(2013)\citenamefont{Liu, Ye, and
  Qi}}]{LiuYeQi}
\bibinfo{author}{\bibfnamefont{C.-X.} \bibnamefont{Liu}},
  \bibinfo{author}{\bibfnamefont{P.}~\bibnamefont{Ye}}, \bibnamefont{and}
  \bibinfo{author}{\bibfnamefont{X.-L.} \bibnamefont{Qi}},
  \bibinfo{journal}{Phys. Rev. B} \textbf{\bibinfo{volume}{87}},
  \bibinfo{pages}{235306} (\bibinfo{year}{2013}).

\bibitem[{\citenamefont{Stone and Lopes}(2016)}]{StoneLopes}
\bibinfo{author}{\bibfnamefont{M.}~\bibnamefont{Stone}} \bibnamefont{and}
  \bibinfo{author}{\bibfnamefont{P.~L. e.~S.} \bibnamefont{Lopes}},
  \bibinfo{journal}{Phys. Rev. B} \textbf{\bibinfo{volume}{93}},
  \bibinfo{pages}{174501} (\bibinfo{year}{2016}).

\bibitem[{\citenamefont{Jackiw and Pi}(2007)}]{JackiwPi}
\bibinfo{author}{\bibfnamefont{R.}~\bibnamefont{Jackiw}} \bibnamefont{and}
  \bibinfo{author}{\bibfnamefont{S.-Y.} \bibnamefont{Pi}},
  \bibinfo{journal}{Phys. Rev. Lett.} \textbf{\bibinfo{volume}{98}},
  \bibinfo{pages}{266402} (\bibinfo{year}{2007}).

\bibitem[{\citenamefont{Teo and Kane}(2010)}]{TeoKane}
\bibinfo{author}{\bibfnamefont{J.~C.~Y.} \bibnamefont{Teo}} \bibnamefont{and}
  \bibinfo{author}{\bibfnamefont{C.~L.} \bibnamefont{Kane}},
  \bibinfo{journal}{Phys. Rev. B} \textbf{\bibinfo{volume}{82}},
  \bibinfo{pages}{115120} (\bibinfo{year}{2010}).

\end{thebibliography}

\appendix
\setcounter{equation}{0}
\makeatletter
\renewcommand{\theequation}{A\arabic{equation}}

\section{Appendix: Vortex and edge mode wavefunctions}

Here we solve for the wavefunctions of the $n$ vortex and edge mode solutions of the Hamiltonian~\eqref{Hamiltonian matrix} in the presence of $\Delta(\b{r}) = \Delta_0 e^{i n \theta}$, now labelling the spinor basis more generally as $\Psi^{\dagger} = (\psi_1, \psi_2, \psi_3, \psi_4)$. We will begin with $p_z = 0$, in which case all modes lie at zero energy. Further, we note that the Hamiltonian in the presence of an axial gauge field can be obtained from the Hamiltonian with $\bm{A}_5 = 0$ via the transformation
\begin{equation}\label{A5 transform}
\mathcal{H}_{\bm{A}_5} = e^{ \alpha^3 \frac{1}{\nabla^2} b} \, \mathcal{H}_{\bm{A}_5 = 0} \, e^{ \alpha^3 \frac{1}{\nabla^2} b} 
\end{equation}
where $\alpha^3$ is the $4 \times 4$ matrix $\alpha^3 = \sigma_3 \otimes \tau_3$, and $b = \epsilon_{ij} \partial_i A_{5,j}$~\cite{JackiwPi}. In light of this, we will solve for the vortex and edge modes with $\bm{A}_5 = 0$, from which the solutions in the presence of $\bm{A}_5(\b{r})$ can be obtained using 
\begin{equation}
\Psi_{\bm{A}_5}(\b{r}) = e^{-\alpha^3 \frac{1}{\nabla^2} b(\b{r})} \Psi_{\bm{A}_5 = 0}(\b{r}). 
\end{equation}
For the axial gauge field vortex~\eqref{axial vortex}, we have $b(\b{r}) = n \pi \delta^{(2)}(x,y)$ and $\frac{1}{\nabla^2} b(\b{r}) = \frac{n}{2} \ln(r)$ and thus $\Psi_{\bm{A}_5}(\b{r}) = r^{- \frac{n}{2} \alpha^3} \Psi_{\bm{A}_5 = 0}(\b{r})$. 

With these conditions, the matrix equation $\mathcal{H}(\b{r}) \Psi(\b{r}) = 0$ becomes
\begin{equation}\label{zero mode diff eqs}
\begin{split}
e^{-i\theta}\bigg(\partial_r - \frac{i}{r} \partial_{\theta}\bigg) \psi_2(\b{r}) + i e^{i n \theta} \, \Delta_0 \, \psi_3(\b{r}) &= 0 \\
 i e^{-i n \theta} \, \Delta_0 \, \psi_2(\b{r}) - e^{i\theta}\bigg(\partial_r + \frac{i}{r} \partial_{\theta}\bigg) \psi_3(\b{r}) &= 0 \\
e^{i\theta}\bigg(\partial_r + \frac{i}{r} \partial_{\theta}\bigg) \psi_1(\b{r}) + i e^{i n \theta} \, \Delta_0 \, \psi_4(\b{r}) &= 0 \\
 i e^{-i n \theta} \, \Delta_0 \, \psi_1(\b{r}) - e^{-i\theta}\bigg(\partial_r - \frac{i}{r} \partial_{\theta}\bigg) \psi_4(\b{r}) &= 0. \\
\end{split}
\end{equation}
Anticipating from the index theorem that the vortex modes will have only $\psi_1, \psi_4 \neq 0$, we seek to solve the last two equations. To do so, we define $\tilde{\psi}_{1,p}$ and $\tilde{\psi}_{4,p}$ by $\psi_1 \equiv r^{n/2} e^{i(p-1)\theta} \, \tilde{\psi}_{1,p}$ and $\psi_4 \equiv -i r^{n/2} e^{-i(n-p)\theta} \, \tilde{\psi}_{4,p} $ for $p = 1, 2, ..., n$ and assume no further $\theta$-dependence. With this, the equations for $\tilde{\psi}_{2,p}(r)$ and $\tilde{\psi}_{3,p}(r)$ read
\begin{equation}\label{high vorticity vortex mode diff eqs}
\begin{split}
\bigg( \partial_r  + \frac{\frac{n}{2}-p+1}{r} \bigg) \tilde{\psi}_{1,p}(r) +  \, \Delta_0 \, \tilde{\psi}_{4,p}(r) &= 0 \\
\Delta_0 \, \tilde{\psi}_{1,p}(r) + \bigg( \partial_r  - \frac{\frac{n}{2}-p}{r} \bigg) \tilde{\psi}_{4,p}(r) &= 0. \\
\end{split}
\end{equation}
These can be solved explicitly when $\Delta_0(r) = \Delta_0$, in which case the solutions read $\tilde{\psi}_{2,p}(r) = K_{\frac{n}{2} - p + 1}(\Delta_0 r)$ and $\tilde{\psi}_{3,p}(r) = K_{\frac{n}{2} - p}(\Delta_0 r)$, where $K_{\alpha}(r)$ is the modified Bessel function of the second kind of order $\alpha$. The full solutions for the $n$ zero modes indexed by $p$ are thus
\begin{equation}\label{higher vorticity wavefunctions}
\begin{split}
\psi_{1,p}(\b{r}) &=  e^{i\pi/4} \,  e^{i(p-1)\theta} \, (\Delta_0 r)^{n/2} \, K_{\frac{n}{2} - p + 1}(\Delta_0 r) \\
\psi_{4,p}(\b{r}) &=  e^{-i\pi/4} \, e^{-i(n-p)\theta} \, (\Delta_0 r)^{n/2} \, K_{\frac{n}{2} - p}(\Delta_0 r). \\
\end{split}
\end{equation}
Note that for $n=p=1$ we have $K_{\frac{n}{2}-p+1}(x) = K_{\frac{n}{2}-p}(x) = K_{\frac{1}{2}}(x) \sim \sqrt{\frac{1}{x}} e^{-x} $ and we recover the previous solution~\eqref{vortex mode}. In the presence of a nonvanishing axial gauge field in the vortex configuration~\eqref{axial vortex} and nonzero $p_z$, we have
\begin{equation}\label{higher vorticity wavefunctions A5}
\begin{split}
\psi_{1,p}(\b{r}) &= e^{i p_z z}  e^{i\pi/4} \,  e^{i(p-1)\theta}  \, K_{\frac{n}{2} - p + 1}(\Delta_0 r) \\
\psi_{4,p}(\b{r}) &= e^{i p_z z}  e^{-i\pi/4} \, e^{-i(n-p)\theta}  \, K_{\frac{n}{2} - p}(\Delta_0 r), \\
\end{split}
\end{equation}
with energy $E = p_z$. 

We now solve for the edge modes, for which $\psi_2$ and $\psi_3$ are nonzero and satisfy the first two of Eqs.~\eqref{zero mode diff eqs} when $p_z$ and $\bm{A}_5$ are set to zero. Similar to before, we assume $\Delta_0(r) = \Delta_0$ and redefine $\psi_2 \equiv r^{-n/2} e^{i(n+p)\theta} \, \tilde{\psi}_{2,p}$ and $\psi_3 \equiv -i r^{-n/2} e^{i(p-1)\theta} \, \tilde{\psi}_{3,p} $. With these assumptions, the equations become 
\begin{equation}\label{high vorticity edge mode diff eqs}
\begin{split}
\bigg( \partial_r  + \frac{\frac{n}{2}+p}{r} \bigg) \tilde{\psi}_2(r) +  \, \Delta_0 \, \tilde{\psi}_3(r) &= 0 \\
\Delta_0 \, \tilde{\psi}_2(r) + \bigg( \partial_r  - \frac{\frac{n}{2}+p-1}{r} \bigg) \tilde{\psi}_3(r) &= 0. \\
\end{split}
\end{equation}
The solutions now involve modified Bessel functions of the second kind, and read
\begin{equation}\label{higher vorticity wavefunctions}
\begin{split}
\psi_{2,p}(\b{r}) &=  e^{i\pi/4} \,  e^{i(n+p)\theta} \, (\Delta_0 r)^{-n/2} \, I_{\frac{n}{2} + p}(\Delta_0 r) \\
\psi_{3,p}(\b{r}) &=  e^{-i\pi/4} \, e^{i(p-1)\theta} \, (\Delta_0 r)^{-n/2} \, I_{\frac{n}{2} + p - 1}(\Delta_0 r) \\
\end{split}
\end{equation}
within in the Weyl material, and $0$ outside of the material. We note that the $x$ and $y$ current for these solutions vanishes everywhere, and thus the discontinuity at the boundary of the Weyl material is allowed. In the presence of nonzero $p_z$ and $\bm{A}_5$, we have
\begin{equation}\label{higher vorticity wavefunctions A5}
\begin{split}
\psi_{2,p}(\b{r}) &=  e^{i p_z z} e^{i\pi/4} \,  e^{-i(p-1)\theta} \,  I_{\frac{n}{2} + p}(\Delta_0 r) \\
\psi_{3,p}(\b{r}) &=  e^{i p_z z} e^{-i\pi/4} \, e^{i(n-p)\theta}  \, I_{\frac{n}{2} + p - 1}(\Delta_0 r), \\
\end{split}
\end{equation}
with energy $E = -p_z$. 

Finally, we note that there also exists an exponentially growing solution of Eqs.~\eqref{high vorticity vortex mode diff eqs} and an exponentially decaying solution of Eqs.~\eqref{high vorticity edge mode diff eqs}, owing to the fact that we have two coupled first order differential equations, and thus two linearly independent solutions. These solutions are not protected by an index theorem however, and will thus couple to bulk modes when we include effects outside of the low energy theory.

\end{document}